\documentclass[11pt]{article}
\usepackage[margin=1in]{geometry}
\usepackage{setspace}
\usepackage{lineno}
\usepackage{graphicx}
\usepackage{booktabs}
\usepackage{caption}
\usepackage[hidelinks]{hyperref}
\usepackage{authblk}
\usepackage[
backend=biber,
style=authoryear,
sorting=nyt,
giveninits=true,
maxcitenames=1,  
mincitenames=1,
maxbibnames=99,  
doi=true,
url=false,
isbn=false
]{biblatex}

\DeclareFieldFormat{doi}{\url{https://doi.org/#1}}
\usepackage{url}

\title{\textbf{Energy per base pair model from NN parameters and its applications in genomic research}}

\author[1]{Shwe Sin Oo\thanks{Corresponding author: E-mail: \texttt{shwe.oo@usm.edu}.}}
\author[1]{Khin Maung Maung\thanks{Corresponding author: E-mail: \texttt{khin.maung@usm.edu}.}}

\affil[1]{School of Mathematics and Natural Sciences, The University of Southern Mississippi, Hattiesburg, MS 39406, U.S.A.}

\begin{document}	

\noindent
\maketitle


\section*{Abstract}
Nearest-neighbor(NN) free energy parameters for DNA are well studied and reliable values of these parameters exist in the literature.
They have been found to be very useful in studying DNA melting and DNA stabilization studies. In this paper, using these parameters, we have constructed a model in which one can define the energy of a base pair between any two neighboring base pairs. This model allows us to use the Boltzmann weighting factor to perform Monte Carlo sampling to probe the average energy per base pair in random sequences that must have existed at the very beginning before life existed. We then employed our model to the publicly available human Genome data. We calculated the average energy per base pair in inter-gene regions where there is no overlap of genes and also for exons and introns separately. We found that (1)  these 'ancient' random sequences still persist in human genome although very scarce (2) the average energy per base pair has drifted about $22.5 \%$ of the ancient value. (3) Our model together with known mutation rates provide us with a new way to look at the 'age of life'.

\noindent\textbf{Key words:} DNA stacking energy, nearest-neighbor model, Monte Carlo simulation, genome evolution, sequence energetics

\section*{Significance}
We provide a simple model for average energy per base pair. This model can be easily employed in Monte Carlo sampling, study of DNA energy landscape and mutation time calculations and in the study of the drift of energy per base pair through evolution. This provides us with a new quantitative measure for genomic studies. This can be used as a new physics based input in bioinformatics research where the main tools are sequence matching, statistical analysis and machine learning.  

\section*{Introduction}

In a game of craps, two dice are rolled and the outcomes are the sum of the two numbers shown on the dice. There are eleven possible outcomes from 2 to 12. In this game, 7 is the most probable outcome because it can be achieved in six different ways; the combinations of the dice are (1,6), (6,1), (2,5), (5,2), (3,4), and (4,3). The next most probable outcomes are 6 and 8 with each outcome having five different combinations. The least probable outcomes are 2 and 12 each having only one combination each to produce them. In the language of statistical mechanics, specific configurations of the two dice such as (2,5) or (1,3) are called microstates. The collections of microstates with the same outcomes, are called macrostates. For example, the microstates (1,3), (3,1) and (2,2) all belong to the macrostate with outcome 4. The number of microstates in a macrostate is called its multiplicity. All microstates are equally probable, but macrostates are not.  Macrostates with higher number of microstates are more probable.

Suppose that we have an unlimited pair of dice available to us. Next, we put each pair into transparent cuboids where the dice can easily roll around. We then throw these cuboids onto a vast land as a single layer in a random fashion. Obviously, the most abundant number of cuboids will show an outcome of 7 and the next most abundant outcome will be 6 and 8. After that there will be 5 and 9 etc. Next, imagine that, over time, some of these cuboids begin to attach to each other in a length wise manner and make cuboid chains of various lengths. For any given length of these chains, the most abundant number of chains will have the average outcome per cuboid close to 7. There will be other chains with average outcome per cuboid less than 7, but these will be less abundant. Now with this dice problem in mind, we consider a DNA sequencing problem.

Suppose we have an unlimited supply of nucleotides in some medium and let them freely attach to each other and form N base pairs long sequences.  Formation of these chains is 'random' apart from the energy requirements in attaching new nucleotides to an already existing chain.  For example, the energy involved in stacking an AT base pair on top of a GC base pair is different from stacking a GC base pair on another GC base pair etc. Therefore, not all arrangements of base pairs will be created with equal frequency. All possible sequences will be formed, but energetically favorable sequences will be more abundant than others.

We will call these sequences, “ancient sequences”. This group of sequences form a macrostate analogous to the macrostate with outcome 7 in the case of the dice problem. 

In this study, we will be concerned with DNA sequences only. The questions we want to ask are (1) In the case of DNA, for a given number of base pairs, what are the most probable ancient sequences and what are their properties? What is the quantitative measure that we can use to classify these sequences? (2) Do the present-day life forms, including us, still possess these ancient sequences? If we do, what would be the longest length that we can find?

Over time, longer sequences are formed, mutations happen and somehow life has started. In our dice analogy, mutations would be equivalent to dice in random cuboids flipping spontaneously and this would change the average outcome per cuboid. Because of mutations, original sequences started to change and evolution gained momentum. Therefore, the next question we want to ask is, (3) How much of the sequences of our DNA has changed from the original ancient sequences and what would be the reasonable measure for this? i.e. how would we quantify these changes? In this paper, we try to answer these questions.

\section*{Stacking Energy and Single Base Pair Energy Model}

In the case of the dice problem, we use the total number shown by the two dice to classify the macro-states. In the case of DNA, there is something called Gibb’s free energy change $\Delta G_{\rm Tot}$. This energy change is given by the sum of stacking energies $\Delta G_{\rm Stk}$,  initiation energies $\Delta G_{\rm init}$, and symmetry energy $\Delta G_{\rm sym}$. Stacking energy is the energy required to stack one Watson-Crick base pair on top of another. Initiation energy accounts for the differences in energy among sequences starting or ending with AT
or CG pairs. The symmetry energy $\Delta G_{\rm sym}$ is the energy parameter associated with self-complementary sequences.  These parameters are well documented in the literature for DNA \parencite{SantaLucia1998}, \parencite{Breslauer1986}, \parencite{Sugimoto1996}, \parencite{Bommarito2000}. Computed parameters for RNA also exist  \parencite{Brown2015}. The following type of study can be done for both DNA and RNA, but for the present work, we concentrate on DNA only.  

The expression for the total free energy change is given in \parencite{SantaLucia1998} as

\begin{equation}
	\Delta G_{\rm Tot} = \sum\limits_i n_i \Delta G_{\rm Stk}(i)   +  \Delta G_{\rm init} + \Delta G_{\rm sym}			
\end{equation}

In the above equation $\Delta G_{\rm Stk}(i)$ are the stacking energies for the 10 Watson-Crick nearest-neighbors (NN). We will use the convention used by (\cite{SantaLucia1998} ). Therefore GT/CA is a dimer with 
5’ G T 3’ and 3’ C A 5’. Then the energy of stacking TA on GC (going from 5’ to 3’) is denoted by $\Delta G (\rm{GT/CA)}$. For $i=1$ to $i=10$,
they are $\Delta G ({\rm AA/TT})$, $\Delta G ({\rm AT/TA})$, $\Delta G {\rm (TA/AT)}$,  $\Delta G \rm{(CA/GT)}$, $\Delta G \rm{(GT/CA)}$, $\Delta G \rm{(CT/GA)}$, $\Delta G \rm{(GA/CT)}$, $\Delta G \rm{(CG/GC)}$, $\Delta G \rm{(GC/CG)}$, $\Delta G \rm{(GG/CC)}$. The numbers $n_i$ are the number of times the same type of stacking happens in the sequence.  The last term is zero if the sequence is not self-complementary. 

Equation (1) is correct and complete, but it is not in a form convenient to define single base pair energy.  Now consider a short sequence of DNA.

\begin{eqnarray*}
	&5'& A~C~T~C~T~A~G~C ~~3'\\
	&3'& T~G~A~G~A~T~C~G~~5'
\end{eqnarray*}

Starting from the left, we will name the first base pair (AT)$=b_0$ , the second base pair (CG)$=b_1$, the third base pair (TA)$=b_2$ etc.

Therefore, an N+2 base pair long sequence can be written as 5’ $b_0~ b_1~ b_2~ \cdots \cdots ~b_N ~b_{N+1}$ 3’. Here the $b_i$'s are the base pairs. Then the total free energy change $\Delta G_{\rm Tot}$ can be written as 

\begin{equation}
\Delta G_{\rm Tot} = \sum\limits_{i=0}^{i=N} \Delta G_{\rm Stk}( b_i \Leftarrow  b_{i+1}) + \Delta G_{\rm init}(b_0)+\Delta G_{\rm init}(b_{N+1} ) +\Delta G_{\rm sym}
\end{equation}

\noindent
The notation is that $\Delta G_{\rm Stk} (b_i \Leftarrow b_{i+1})$ means the stacking energy to stack the base pair $b_{i+1}$ on $b_i$. The orientation is 5' $b_i~b_{i+1}$ 3'. The terms $\Delta G_{\rm init}(b_0)$ and $\Delta G_{\rm init}(b_{N+1} )$ are the initiation energies for the first and the last base pair and $\Delta G_{\rm sym}$ is the energy for symmetry which is non-zero only for self-complementary sequences. We now rearrange equation (2) and obtain

\begin{equation}
	\Delta G_{\rm Tot}=\sum\limits_{i=1}^{i=N} \frac{( \Delta G_{\rm Stk}( b_{i-1}\Leftarrow b_i) + \Delta G_{\rm Stk}( b_i\Leftarrow b_{i+1} ) )}{2} + \Delta G_{\rm edge}
\end{equation}

\begin{equation}
\Delta G_{\rm edge}= \frac{1}{2}\Delta G_{\rm Stk}(b_0 \Leftarrow b_1)+\frac{1}{2} \Delta G_{\rm Stk}(b_N \Leftarrow b_{N+1}) +\Delta G_{\rm init}(b_0)+\Delta G_{\rm init}(b_{N+1} ) +\Delta G_{\rm sym}
\end{equation}

We now define the energy per base pair for i=1 to i=N as

\begin{equation}
E(b_i) = \frac{( \Delta G_{\rm Stk}( b_{i-1}\Leftarrow b_i) + \Delta G_{\rm Stk}( b_i \Leftarrow b_{i+1} ) )}{2} + \frac{1}{N}\Delta G_{\rm edge}
\end{equation}

It is easy to check that for an N+2 base pair long sequence (i=0 to N),

\begin{equation}
\Delta G_{\rm Tot}=\sum\limits_{i=1}^{i=N} E(b_i)
\end{equation}

This way of rewriting the $\Delta G_{\rm Tot}$ let us define single base pair energy in terms of stacking energy. $\Delta G_{\rm edge}$ is obviously small. The stacking energies on the average for a dimer contributes about -1.56 kcal/mol and the initiation energies contribute about +1 kcal/mol and the energy from symmetry is no more than +0.5 kcal/mol \parencite{SantaLucia1998}. Therefore, regardless of the length of the sequence, in total, $\Delta G_{\rm edge}$ contributes on the average, 0.44 kcal/mol for sequences without symmetry and about 0.94 kcal/mol for self-complementary sequences. Therefore large N sequences, we can safely neglect these terms. In this work we use Table I from \parencite{SantaLucia1998}. This concept of single base pair energy is useful for Monte Carlo sampling, study of DNA energy landscape, and also in studies of mutation processes.

\section*{ Ancient sequences}
In order to find the most probable sequences for a given number of base pair N, we need to find all possible sequences and calculate their energies. Then by binning these sequences according to their energies, we can classify them into macrostates. Of course, binning depends on our choice of bin width.  Obviously, if the bin width is small enough, sequences with similar energies will fall into the same bin. Then, the most abundant sequences will be in the tallest bin.  In an N base pair long sequence, the number of possible sequences is $4^N$. Therefore even for a modest value of N=25, the number is $1.25\times 10^{15}$, which is incredibly large. Therefore, instead of finding all possible sequences, we use Mote Carlo sampling method \parencite{Metropolis1953}, \parencite{Hastings1970}). For N=18, 19, 20, 21, and 22. We find the sequences by Monte Carlo sampling and calculate the average energy per base pair of each sequence. For each N we used 1 million, 3 million, 5 million up to 45 million Monte Carlo steps.  For each N, we put the sequences in bins according to their average energy per base pair and fit the resulting histograms to  Gaussians.  In Fig (1), we show the results for N=21. In each bin, the sequences are selected according to their energies. Therefore, the tallest bin will contain the most number of sequences, whose average energies per base pair are inside the bounds $ E_c \pm \Delta E /2$, where $E_c$ is the energy representing the center of the bin and $\Delta E$ is the width of the bin. $E_c$, the mean value of average energy per base pair reaches convergence at 5 million Monte Carlo steps. For each N, we extract the sequences from the tallest bin and tried to locate them in human genome. As N becomes larger, it becomes more scarce to find. We can understand this from the following formula:

\begin{equation}
	P =1-exp(-(M-N+1)( \frac{1}{4})^N))
\end{equation}

Here, $P$ is the probability of finding one match, $M$ is the number of base pairs in human genome, $N$ is the number of base pairs in our sequence that we want to find in human genome. In our example, taking $3\times 10^9$ base pairs for M, and N=21, we obtain approximately $P\equiv 0.00068$. The longest 'ancient sequence' we have found in human genome \parencite{Yates2020Ensembl}  is for N=21 and it is 'CAGGCAGGGGTTGGGGGGCCA'  which is located in chromosome { 5 } at location  12159488 to 12159508. This sequence was obtained from the Monte Carlo run with 45 million steps. We found that $E_c=1.78$ kcal/mol for N=21. This is the typical average energy per base pair for a 21 base pair long ancient sequence.

\begin{figure}[htbp]
	\centering
	\includegraphics[width=0.7\textwidth]{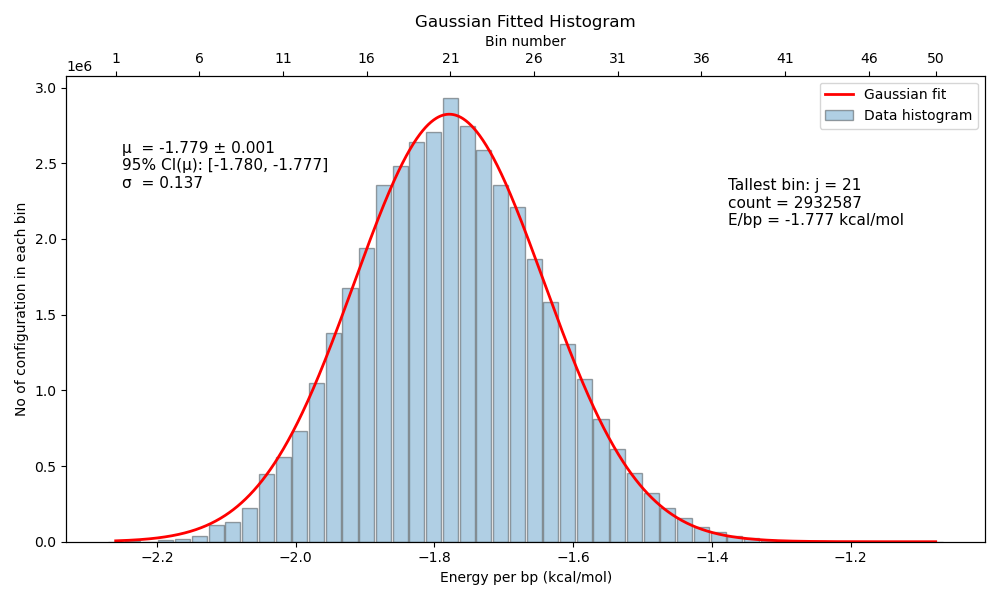}
	\caption {Energy distribution from Monte Carlo simulation.}
             {Average energy per base pair $\mu =-1.78$ kcal/mol from the fit.} 
	\label{fig:energy}
\end{figure}

\section*{The drift of the average energy per base pair}

Next we investigate what the average energy per base pair for the human genome is. Again, we use the publicly available human genome data source  \parencite{Yates2020Ensembl}. We performed this study for three different categories of sequences. The first category is the  sequences from intergenic regions which are clear of any nested gene structure. The second is from exons inside genes and the third is from the introns inside genes. 

To get the intergenic regions, for a given chromosome, we first find the gene names and their locations. Next, we identified the regions which are between two genes and not also enveloped inside a gene. These regions are truly intergenic and do not contain any coding regions. Now, to be consistent with the procedure that we have used for the 'ancient' sequences, we read these intergenic regions with a window that is 21 base pair long and find the average energy per base pair for each window. After a window is read, and the average energy per base pair calculated, we move the window by one base pair and repeat the procedure again. The maximum number of times a full size window can be read is $M-(w-1)$.
Here $M$ is the total number of base pairs in the region, and $w$ is the window size. See Fig(2) and Fig(3) where we have shown the histograms from two different intergenic regions from two different chromosomes. 

\begin{figure}[htbp]
	\centering
	\includegraphics[width=0.7\textwidth]{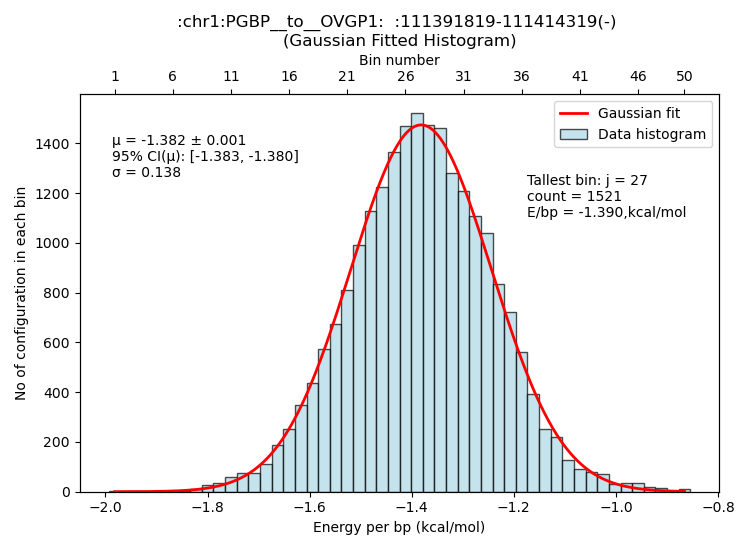}
	\caption{Intergenic region between genes PGBP1 and OVGP1 in chromosome 1}
	{Average energy per base pair $\mu =-1.382$ kcal/mol from the fit.} 
	\label{fig:energy}
\end{figure}

\begin{figure}[htbp]
	\centering
	\includegraphics[width=0.7\textwidth]{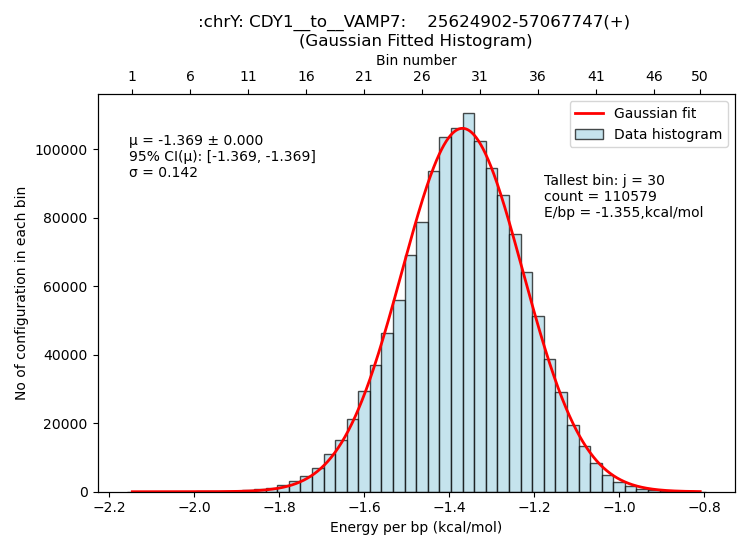}
	\caption{Intergenic region between genes CDY1 and VAMP7 in chromosome Y}
	{Average energy per base pair $\mu =-1.369$ kcal/mol from the fit.} 
	\label{fig:energy}
\end{figure}
To study the exons and introns, in each chromosome, we picked some genes arbitrarily and separated them into exons and introns regions. Then we merge all exons together that belong to the same Ensemble Transcript ID. We did the same for introns. We then read this combined exons using a moving window of 21 base pairs, calculate the energies and bin them. We did the same calculations for combined introns. In Fig(4) and Fig(5) we show the results from an exon and an intron. Similar results were obtained for other Ensemble Transcription IDs. In the following, we give the summary of our findings for intergenic region, exons and introns,  
\begin{figure}[htbp]
	\centering
	\includegraphics[width=0.7\textwidth]{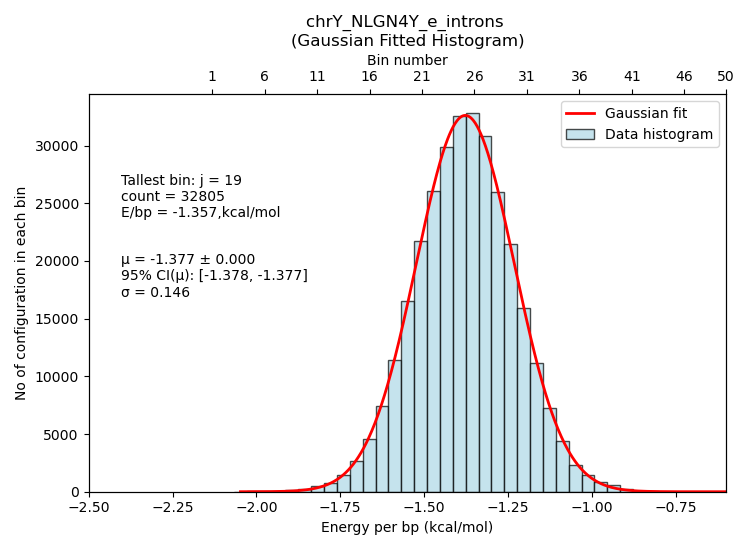}
	\caption{Introns from gene NLGN4Y-e from chromosome Y}
	{Average energy per base pair $\mu =-1.377$ kcal/mol from the fit.} 
	\label{fig:energy}
\end{figure}

\begin{figure}[htbp]
	\centering
	\includegraphics[width=0.7\textwidth]{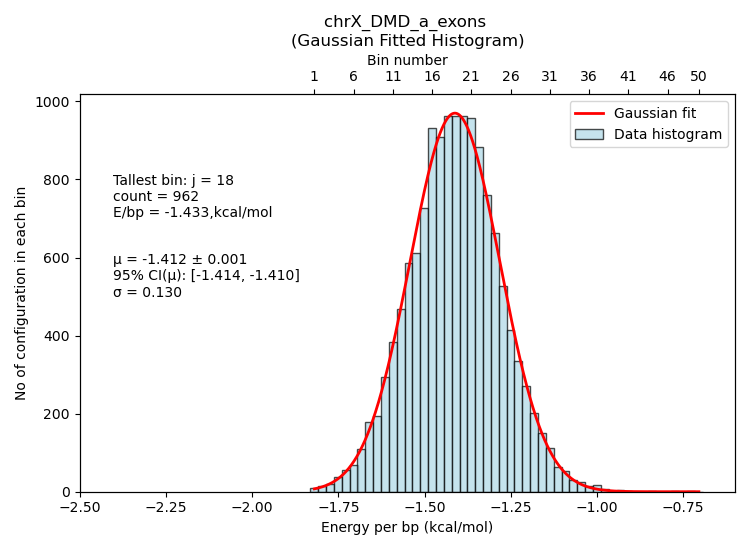}
	\caption{Intergenic Exons from gene DMD-a from chromosome X}
	{Average energy per base pair $\mu =-1.412$ kcal/mol from the fit.} 
	\label{fig:energy}
\end{figure}

We find that the average energy per base pair for the ancient sequences is 1.78 kcal/mol. For the current human DNA, from five chromosomes, (chromosome 1, chromosome 5, chromosome 14, chromosome X and chromosome Y), we picked  10 segments of intergenic regions from each chromosome, totaling 50 segments. To study exons and introns inside genes, we first picked five genes each from the five chromosomes. Then in each gene, we selected five alternative splicing of exons and five alternative splicing of introns.   We show two such cases in Fig (4) and Fig (5). Their gaussian fitted $\mu$ values do not vary too much from their average $1.38$ kcal/mol. By comparing this average $1.38$ kcal/mol to the ancient average $1.78$ kcal/mol, we see that the drift of energy per base pair is 0.4 kcal/mol which is 22.5$\%$ of the ancient sequence value. It is reasonable to assert that there was no life when the random DNA sequences were abundant. Because of mutations caused by many different conditions, the average energies started shifting and at some point, order set in, and life started.  As far as we are aware, this is the first time a quantitative measure has been found that distinguishes between non-life and life. 
 
\section*{Estimation of the age of 'life'}

We assume that when RNA or DNA sequences were forming at the beginning, the process must include random attachment of available nucleotides obeying energy requirements. Therefore, if we look at any N base pair long sequences, the most abundant sequences will be the energetically favored ones. As mentioned above, the average energy per base pair of these 'ancient sequences' is 1.78kcal/mol. We assume that there was no life present at that time. In time, the sequences might link to one another, attached new base pairs, and became longer. At the same time, mutation started taking place and the sequences started to change. Eventually, at some point, life started. Now, using current human genome data 
 \parencite{Yates2020Ensembl}    we have found that the current average base pair energy (based on 21 base pair long sequence ) is 1.38 kcal/mol. 

In order to estimate the age of life, we started by generating ancient sequences in large amounts; 1 million, 3 million, 5 million up to 45 million. For a given batch, we collected the sequences that produce 1.76 kcal/mol  to 1.78 kcal/mol average energy per base pair. Then we sifted out only the unique sequences. For each sequence, we mutate one base pair at a time and calculate the average energy per base pair. We record the number of mutations needed when the average energy per base pair reaches the current day values $-1.38$ kcal/mol. When all sequences are done, we then calculate the average number of mutations. We do not count the sequences that did not successfully reach the current day average energy per base pair. We find that the average number of mutations converges to 15.877 when the number of starting sequences are 5 million. 

In order to calculate the age of life, we start with the definition of mutation rate.  We are not performing experiment with diploid genome but just started with
one sequence. Therefore if $N$ is a sequence of base pairs and $M$ is the number of mutations during the observation time $T$ in years, then mutation rate $R$, is
\begin{equation}
	R=\frac{M}{NT}~~~~{\rm or~for~ovservation~time~T,~}~~~~~~~T=\frac{M}{NR}
\end{equation}
In our case, we use $M=15.877$, $N=21$ and $R$ from the given references. Some authors give the mutation rate as number of mutations per base pair per generation, where one generation equals thirty 30 years.  

According to \parencite{Kong2012Denovo} the mutation rate is $1.20 \times 10^{-8}$ per
generation and from \parencite{Rahbari2016}  it is $1.28 \times 10^{-8}$ per nucleotide per generation. From \parencite{Josson2017}  the mutation rate is $1.29\times 10^{-8}$ per nucleotide per generation and
$4.27 \times 10^{-10} $ per base pair per year. 
\vskip 30pt

\begin{tabular}{|c|c|c|}
	\hline
	\textbf{Reference} & \textbf{Mutations per bp per generation} & \textbf{Observation time in years} \\
	\hline
	 \parencite{Jarad2010} & $1.1 \times 10^{-8}$  & $2.1 \times 10^9$  \\
	\hline
	\parencite{Kong2012Denovo} & $1.2 \times 10^{-8}$ & $1.89 \times 10^{9}$ \\
	\hline
	\parencite{Rahbari2016} & $1.28 \times 10^{-8}$ & $1.77 \times 10^{9}$ \\
	\hline
	\parencite{Josson2017} & $1.29\times 10^{-8}$& $1.75 \times 10^{9}$\\
	\hline
\end{tabular}

\vskip 20pt
This demonstrate how to estimate the age of life using our method. This method can also be used to find the lapse of time between two segments of equal length DNA by mutation. 

\section*{Conclusion and Discussion}
We have presented a method to define energy of a single base pair inside a DNA sequence. Using this method, it is easy to employ Mote Carlo methods such as Metropolis-Hastings algorithm. We found that if DNA sequences are formed randomly but obeying the energy requirements, the average energy per base pair is (based on 21 bp) is $1.78$ kcal/mol. We found that it is very scarce to find these ancient sequences in modern human genome. We did find a sequence of 21 base pair long in chromosome 5. Next, we find that the average energy per base pair for intergenic regions, exons and introns is on the average $1.38$ kcal/mol. Therefore, we conclude that there is a drift of the average energy. Somewhere between $1.78$ kcal/mol and $1.38$kcal/mol base pair energies, life started. As far as we are aware, this is the first time a quantitative measure has been introduced related to transition from no-life to life. Using our method, we also calculated the 'age of life' using known mutation rate in the literature. Although it might seem on the lower side of other estimates of the age of life using, different methods, we believe that our method is robust and should be tested with different genomes.

\printbibliography[title={Literature Cited}]

@article{SantaLucia1998,
  author  = {SantaLucia Jr., J.},
  title   = {A unified view of polymer, dumbbell, and oligonucleotide DNA nearest-neighbor thermodynamics},
  journal = {Proc Natl Acad Sci U S A.},
  volume  = {95},
  number  = {4},
  pages   = {1460--1465},
  year    = {1998},
  doi     = {10.1073/pnas.95.4.1460}
}

@article{Sugimoto1996,
  author  = {Sugimoto, Naoki et al.},
  title   = {Improved Thermodynamic Parameters and Helix Initiation Factor to Predict Stability of DNA Duplexes},
  journal = {Nucleic Acids Research},
  volume  = {24},
  number  = {22},
  pages   = {4501--4505},
  year    = {1996},
  doi     = {10.1093/nar/24.22.4501}
}

@article{Bommarito2000,
  author  = {Bommarito, Salvatore and Peyret, Nicolas and SantaLucia Jr, John},
  title   = {Thermodynamic parameters for DNA sequences with dangling ends},
  journal = {Nucleic Acids Research},
  volume  = {28},
  number  = {9},
  pages   = {1929--1934},
  year    = {2000},
  doi     = {10.1093/nar/28.9.1929}
}

@article{Yates2020Ensembl,
  author  = {Yates, Andrew D. and others},
  title   = {Ensembl 2020},
  journal = {Nucleic Acids Res},
  year    = {2020},
  volume  = {48},
  number  = {D1},
  pages   = {D682--D688},
  doi     = {10.1093/nar/gkz966}
}

@article{Kong2012Denovo,
  author = {Kong, Augustine
            and Frigge, Michael L.
            and Masson, Gisli
            and Besenbacher, Soren
            and Sulem, Patrick
            and Magnusson, Gisli
            and Gudjonsson, Sigurjon A.
            and Sigurdsson, Asgeir
            and Jonasdottir, Aslaug
            and Jonasdottir, Adalbjorg
            and Wong, Wendy S. W.
            and Sigurdsson, Gunnar
            and Walters, G. Bragi
            and Steinberg, Stacy
            and Helgason, Hannes
            and Thorleifsson, Gudmar
            and Gudbjartsson, Daniel F.
            and Helgason, Agnar
            and Magnusson, Olafur Th.
            and Thorsteinsdottir, Unnur
            and Stefansson, Kari},
  title   = {Rate of de novo mutations and the importance of father’s age to disease risk},
  journal = {Nature},
  year    = {2012},
  volume  = {488},
  number  = {7412},
  pages   = {471--475},
  doi     = {10.1038/nature11396}
}

@article{Breslauer1986,
  author  = {Breslauer, K. J.
            and Frank, R.
	    and Blocker, H.
	    and Marky, L A.},
  title   = {Predicting DNA duplex stability from the base sequence},
  journal = {Proc Natl Acad Sci U S A.},
  year    = {1986},
  volume  = {83},
  number  = {11},
  pages   = {3746--3750},
  doi     = {10.1073/pnas.83.11.3746}
}

@article{Rahbari2016,
  author  = {Rahbari, Raheleh and, et al},
  title   = {Timing, rates and spectra of human germline mutation},
  journal = {Nat Genet.},
  volume  = {48},
  number  = {2},
  pages   = {126-133},
  year    = {2016},
  doi     = {10.1038/ng.3469}
  
}

@article{Brown2015,
  author  = {Brown, Reid F. and Andrew, Casey T. Andrews and Elcock, Adrian H.},
  title   = {Stacking free energies of all DNA and RNA nucleoside pairs
  and dinucleoside-monophosphates computed using recently revised AMBER parameters and
  compared with experiment  },
  journal = {J Chem Theory Comput. },
  year    = { 2015},
  volume  = {11 },
  number  = {5 },
  pages   = {2315--2328 },
  doi     = {doi:10.1021/ct501170h }
}

@article{Josson2017,
  author  = {Josson, H. etal },
  title   = { Parental influence on human germline de novo mutations in
1,548 trios from Iceland.},
  journal = {Nature},
  year    = {2017 },
  volume  = { },
  number  = { },
  pages   = { },
  doi     = {doi: 10.1038/nature24018 }
}

@article{Jarad2010,
  author  = {Jarad, C. Roach, et al},
  title   = {Analysis of Genetic Inheritance in a Family Quartet by Whole
Genome Sequencing},
  journal = {Science},
  year    = {2010},
  volume  = {328},
  number  = {5978},
  pages   = {636--639},
  doi     = {10.1126/science.1186802}
}

@article{Metropolis1953,
  author  = {Metropolis, N. and Rosenbluth, A. W. and Rosenbluth, M. N. and Teller, A. H. and Teller, E.},
  title   = {Equation of State Calculations by Fast Computing Machines},
  journal = {Journal of Chemical Physics},
  volume  = {21},
  pages   = {1087--1092},
  year    = {1953},
  doi     = {10.1063/1.1699114}
}

@article{Hastings1970,
  author  = {Hastings, W. K.},
  title   = {Monte Carlo sampling methods using Markov chains and their applications},
  journal = {Biometrika},
  volume  = {57},
  pages   = {97--109},
  year    = {1970},
  doi     = {10.1093/biomet/57.1.97}
}

\end{document}